\pdfoutput=1
\documentclass[sigconf]{acmart}

\AtBeginDocument{%
  \providecommand\BibTeX{{%
    \normalfont B\kern-0.5em{\scshape i\kern-0.25em b}\kern-0.8em\TeX}}}

\renewcommand\footnotetextcopyrightpermission[1]{}

\setcopyright{none} % No copyright notice required for submissions

\usepackage{amsfonts,amsmath}
\usepackage{mathtools}
\usepackage{xcolor}
\usepackage{xspace}
\usepackage{multirow}
\usepackage[ruled,vlined]{algorithm2e}
\usepackage{enumitem}
\usepackage{hyperref}
\usepackage{subcaption}
\usepackage{color, colortbl}
\definecolor{Gray}{gray}{0.9}
\definecolor{CRed}{RGB}{255 204 204}
\definecolor{CGreen}{RGB}{221 255 153}
\usepackage{bm}
\definecolor{color1}{HTML}{1F77B4}
\definecolor{color2}{HTML}{FF7F0e}
\definecolor{color3}{HTML}{2CA02C}
\definecolor{color4}{HTML}{D62728}
\definecolor{color5}{HTML}{9467BD}
\newcommand{\removetext}[1]{}
 
\DeclareMathOperator*{\argmax}{argmax}
\newcommand{\datasetowner}{$DO$\xspace}
\definecolor{greena}{rgb}{0.0, 0.7, 0.1}
\newcommand{\newtext}[1]{\textcolor{black}{#1}}

\begin{document}

%%\title{Real-time Attacks Against Deep Reinforcement Learning Policies}
\title{On the Effectiveness of Dataset Watermarking in Adversarial Settings}

\author{Buse G.  A.  Tekgul}
\affiliation{%
\institution{Aalto University}
\city{Espoo}
\country{Finland}}
\email{buse.atlitekgul@aalto.fi}

\author{N. Asokan}
\affiliation{%
\institution{University of Waterloo \& Aalto University}
\city{Waterloo}
\country{Canada}}
\email{asokan@acm.org}

\renewcommand{\shortauthors}{Tekgul,  et al.}

%%%%%%%%% ABSTRACT
\begin{abstract}
%In recent years, it has been proven that adversaries can steal the functionality of machine learning (ML) models using the exposed application interfaces. Therefore, protecting the ownership of ML models has received a considerable attention, and numerous model watermarking techniques have been proposed to trace the original owner of the stolen ML models. Unlike the interest in model watermarking, there have been few methods verifying the ownership of datasets used to train and deploy ML models in an unauthorized way. 
In a data-driven world, datasets constitute a significant economic value. 
%Datasets are used for various purposes ranging from statistical analysis to building machine learning (ML) models. 
Dataset owners who spend time and money to collect and curate the data are incentivized to ensure that their datasets are not used in ways that they did not authorize. When such misuse occurs, dataset owners need technical mechanisms for demonstrating their ownership of the dataset in question. \emph{Dataset watermarking} provides one approach for ownership demonstration 
which can, in turn, deter unauthorized use. In this paper, we investigate \newtext{a recently proposed data provenance method,} \emph{radioactive data}~\citep{sablayrolles2020radioactive}, to assess if it can be used to demonstrate \emph{ownership} of (image) datasets used to train machine learning (ML) models. \newtext{The original paper~\citep{sablayrolles2020radioactive} reported that radioactive data is effective in white-box settings.} We show that while this is true for large datasets with many classes, it is not as effective for datasets where the number of classes is low $(\leq 30)$ or the number of samples per class is low $(\leq 500)$. \newtext{We also show that, counter-intuitively, the black-box verification technique described in~\cite{sablayrolles2020radioactive} \emph{is} effective for all datasets used in this paper, even when white-box verification in~\cite{sablayrolles2020radioactive} is not.} Given this observation, we show that the confidence in white-box verification can be improved by using watermarked samples directly during the verification process. We also highlight the need to assess the robustness of radioactive data if it were to be used for ownership demonstration since it is an adversarial setting unlike provenance identification.%, ownership demonstration is an adversarial setting. 

Compared to dataset watermarking, ML model watermarking has been explored more extensively in recent literature. However, most of the state-of-the-art model watermarking techniques can be defeated via model extraction~\cite{shafieinejad2021robustness}. %This has led to recent proposals for ML model ownership demonstration techniques that can withstand model extraction, such as inserting watermarks into model responses to queries~\cite{szyller2019dawn} or via model fingerprinting \cite{lukas2019deep}, although they have known limitations. 
We show that radioactive data can effectively survive model extraction attacks, which raises the possibility that it can be used for \emph{ML model ownership verification} robust against model extraction.

\end{abstract}

%\begin{CCSXML}
%<ccs2012>
%<concept>
%<concept_id>10002978</concept_id>
%<concept_desc>Security and privacy</concept_desc>
%<concept_significance>500</concept_significance>
%</concept>
%<concept>
%<concept_id>10010147.10010257</concept_id>
%<concept_desc>Computing methodologies~Machine learning</concept_desc>
%<concept_significance>500</concept_significance>
%</concept>
%<concept>
%<concept_id>10010147.10010257.10010293.10010294</concept_id>
%<concept_desc>Computing methodologies~Neural networks</concept_desc>
%<concept_significance>500</concept_significance>
%</concept>
%</ccs2012>
%\end{CCSXML}



\maketitle

\pagestyle{plain}

%%%%%%%%% BODY TEXT
%auto-ignore
\section{Introduction}\label{sec:introduction}

Datasets are an essential component in scientific research and have played a major role in the advancement of artificial intelligence and machine learning (ML). The availability of large-scale datasets is critical for obtaining high-quality ML models in different tasks. Big databases such as ImageNet~\cite{deng2009imagenet}, OpenImages~\cite{kuznetsova2018open} and 
Amazon reviews~\cite{mcauley2015image} have enabled a rapid progress in image classification, segmentation, and recommendation tasks. 

As datasets have gained importance and economic value, protecting the rights of the dataset while allowing the sharing of datasets becomes a major issue. There are different methods to prevent unauthorized access to a private dataset, %including blockchain-based encryption mechanisms for data sharing~\cite{chen2019blockchain} 
such as allowing operations on the encrypted data~\cite{martins2017survey}, although these are computationally expensive. 
%or enabling simple operations (e.g., search, sort, query) on encrypted datasets~\cite{egorov2016zerodb}. 
However, there is little prior work verifying the ownership~\cite{li2020open,kim2020digital} or establishing the provenance~\cite{sablayrolles2020radioactive} of a dataset. This is particularly important in the event of \emph{misuse}. 
%publicly shared datasets, when there is a misuse during the deployment of ML models trained via these datasets. 
Examples of misuse include unauthorized monetization of ML models trained from the dataset, using such models outside a designated permitted context (such as a specific geographical area, in the case of models trained from privacy-sensitive data), or violation of other conditions stipulated in this licence under which the data was distributed. Current dataset ownership verification mechanisms rely on \emph{dataset watermarking}: modifying a small subset of the training set and using this subset, or the knowledge carried within, for ownership verification, when this training set is used to build and deploy an ML model. In dataset watermarking, model trainers have access to the complete training set. Therefore, adversaries can detect or even discard watermarked samples from the training set, if these samples can be recognized by a simple visual inspection or any outlier detection technique. Moreover, adversaries can choose any model architecture and optimization technique for training and modify the training samples during input pre-processing. Therefore, an effective dataset watermarking method should be 1) independent from the model architecture and the learning procedure, 2) robust to input transformations, and 3) visually imperceptible. Based on these criteria, dataset watermarking techniques that rely on backdooring~\cite{li2020open} are not feasible, since these methods change the correct label of watermarked samples, usually leave a perceptible artifact on these samples, and the pattern used for the watermarking can be recovered by different methods~\cite{wang2019neural,aiken2021neural}. On the contrary, radioactive data ~\cite{sablayrolles2020radioactive} satisfies these criteria explained above, since it keeps the true labels while adding watermarks to features instead of pixels. Similarly, the data watermarking framework in~\cite{kim2020digital}, which is proposed for the ownership verification of audio classification datasets, embeds the watermark into the frequency component instead of the time-domain representation. In this paper, we focus on image classification datasets and leave the method in~\cite{kim2020digital} as out of scope. 

%In the last decade, systems based on deep neural networks (DNNs) have demonstrated dramatic success, thus providing a business advantage over the competitors. 

Similar to datasets, there has been extensive research for demonstrating the ownership of ML models via model watermarking~\cite{uchida2017embedding, adi2018turning, jia2021entangled}. However, recent studies~\cite{shafieinejad2021robustness, lukas2022watermarkingsok} show that many state-of-the-art model watermarking methods are not robust to model extraction attacks that obtain a surrogate model with the same functionality as the protected model by querying the victim model's prediction API and exploiting responses for each query. This limitation led researchers to propose different ownership verification mechanisms such as dynamic watermarking~\cite{szyller2019dawn}, fingerprinting~\cite{lukas2019deep,zhao2020afa}, and dataset inference~\cite{maini2021dataset}, despite their intrinsic flaws. These methods are proposed to reduce the incentive for model extraction attacks by identifying that the surrogate model is derived from the original model, and validate the legitimate owner of the surrogate model using the knowledge added during the training of the original model.

%Since obtaining DNNs with a high performance is an expensive process, this might motivate adversaries to steal complex DNN models in order to monetize or redistribute them in unauthorized ways without the substantial cost~\cite{orekondy2019knockoff,atli2020extraction}. In order to reduce incentive for such attacks, various ownership verification methods have been developed, including model watermarking~\cite{uchida2017embedding, adi2018turning, szyller2019dawn,jia2021entangled}, fingerprinting~\cite{lukas2019deep,zhao2020afa}, and dataset inference~\cite{maini2021dataset}. These methods identify that the stolen model is derived from the original model, and validate the legitimate owner of the model by leveraging the knowledge added during the training of the original model. %Therefore, it is necessary to develop techniques using which model owners can ensure ownership of their models. %Dataset watermarking can also trace if any DNN trained using a privacy sensitive training set that is deployed outside of its restricted area of use. Another usage of the dataset watermarking could be identifying all models trained using the licensed training set, where the training set is distributed together with licenses specifying a limited quota of ML models that can be trained from it. 

\subsubsection*{Goal and Contributions:} \newtext{We aim to evaluate the effectiveness of using radioactive data~\cite{sablayrolles2020radioactive}, for ownership verification of (image classification) datasets.} %by using different image classification datasets. 
We also explore whether radioactive data has a \emph{transitive property}: Can it be used as an alternative ownership verification technique for models stolen using prediction APIs? Our contributions are as follows:
\begin{enumerate}[labelindent=0pt]
	\item We reproduce radioactive data~\cite{sablayrolles2020radioactive} using different datasets (CIFAR10, CIFAR100~\cite{krizhevsky2009learning} and subsets of CIFAR100). We confirm that black-box verification~\cite{sablayrolles2020radioactive} is effective and generalizable across different datasets. However, the effectiveness of white-box verification algorithm in~\cite{sablayrolles2020radioactive} is limited for datasets having small number of classes $(\leq 30)$ or number of samples per class $(\geq 500)$. \newtext{This is counter-intuitive in adversarial settings,} because white-box verification assumes complete access to the suspected model, whereas black-box verification only assumes API access (Section~\ref{ssec:reproduction}).
	\item  %We claim that there is a discrepancy between white- and black-box verification: Despite the complete access to the DNN model, white-box verification suffers from a low level of confidence. 
	We show that the confidence is increased when watermarked samples are used for white-box verification instead of a held-out test or validation set as in the original paper~\cite{sablayrolles2020radioactive} (Section~\ref{ssec:improvedwhitebox}). 
	\item We show that radioactive data survives model extraction, raising the question of whether it can be used as an alternative ownership verification method against model extraction attacks. %that reproduce the functionality of a victim model using predictions accessed via the victim's API. 
	We show that the knowledge contained in watermarks transfers to surrogate models and the effect of watermarks still remain during model extraction (Section~\ref{sec:againstmodelextraction}).
	%is highly affected by the number of classes in the dataset, watermarking ratio and the software version used to implement it.
	%\item We also observe that the effectiveness of Radioactive-data appears to be brittle: the white-box verification is negatively affected if a recent PyTorch version is used to train models with the watermarked dataset.  
	%\item We propose \emph{\ourmethod}: a watermark removal method that can evade the dataset ownership verification in image classification tasks. \ourmethod allows adversaries to completely circumvent backdoor-based watermarking~\cite{li2020open} with a minor loss in model performance ($\leq 4pp$). It also allows adversaries to evade Radioactive-data, except in the case of classification models involving a large number (e.g., $\geq 100$) of classes.
	%We show that \ourmethod effectively evades different ownership verification techniques with a trivial decrease ($\leq 5pp$) in the model performance in datasets with a small number of classes.   
\end{enumerate}

%auto-ignore
\section{Background and Related Work}\label{sec:background}
\subsection{Deep Neural Networks}\label{ssec:dnn}

DNN is a function $F$ that maps an input $ x \in  \mathbb{R}^n$ into an output $y \in \mathbb{R}^m$, where $n$ is the number of input features and $m$ is the dimension of the output. In classification problems, $m$ refers to the number of classes, and 
$F(x)$ outputs a vector of length $m$ containing probabilities or log-probabilities that $x$ belongs to each class $y_j$, where $j \in \lbrace 1, m\rbrace$. The predicted class can be found by $\hat{F}(x) = \argmax(x)$. $F$ can be divided into two parts: a \emph{feature extractor} $\phi(x): \mathbb{R}^n \to \mathbb{R}^d$ followed by a \emph{linear classifier}, i.e., the last layer of the DNN
$\zeta(\phi(x))$, where $\zeta: \mathbb{R}^d \to \mathbb{R}^m$. %usually $d \geq m$.

DNNs are trained to approximate a perfect oracle function $\hat{F} \sim \mathcal{O}_f$, where $\mathcal{O}_f: \mathbb{R}^n \rightarrow C$ gives the true class $c$ for any sample $x \in \mathbb{R}^n$. How well $\mathcal{O}_f$ is approximated can be evaluated by obtaining a prediction accuracy $Acc(F)$ on an holdout test set of the ground truth data. 

\subsection{Dataset Watermarking}\label{ssec:datasetwm}
Watermarking is the technique of embedding secret information, which belongs to the original owner or designer, into digital media as a means to prove the legitimate ownership. Similarly, dataset watermarking~\cite{sablayrolles2020radioactive,li2020open,kim2020digital} traces the dataset that is used for training ML models back to its original owner \datasetowner. 

Dataset watermarking consists of two phases: \emph{embedding} and \emph{verification}. In the embedding phase, the watermark $wm$, i.e., a special information, is inserted into a sample $x_{wm}$ inside a subset $D_{wm}$ of the dataset $D$. After the embedding, \datasetowner obtains a watermarked dataset $\tilde{D} = (D \setminus D_{wm}) \cup \tilde{D}_{wm}$, where $\tilde{D}_{wm}$ includes modified, watermarked samples $\tilde{x}_{wm}$. In the verification phase, a $\textsc{Verify}$ function is defined to prove the ownership of $\tilde{D}$. Assuming a DNN model $F$ is trained using $\tilde{D}$ to approximate the oracle $\mathcal{O}_f$, a successful ownership verification is expressed in Equation~\ref{eq:tw_dataset}.
%In dataset watermarking~\citep{sablayrolles2020radioactive,li2020open,kim2020digital}, the dataset owner can verify if its dataset $D$ was used in the training of $F$ by obtaining some special information, i.e., the watermark that is embedded into the subset of the dataset and transferred into the model. After the watermark embedding, ownership verification is performed with the $\textsc{Verify}$ function. %A successful embedding and ownership verification can be defined as follows: 
%Let's assume that $D_v$ is a watermarked dataset that has a subset $v$ including watermarked samples, and $wm$ denotes a watermark pattern or an additional knowledge embedded in each sample in $v$. Then, any $F$ trained using $D_v$ to approximate $\mathcal{O}_f$ must transfer some information about $wm$. Conditions for a successful ownership verification are expressed in Equation~\ref{eq:tw_dataset}.
\begin{equation}\label{eq:tw_dataset}
	 ((\tilde{D}, F) \rightarrow \tilde{F} \; ) \wedge (\tilde{F} \sim \mathcal{O}_f) \Rightarrow \textsc{Verify}(\tilde{F}, \tilde{D}_{wm}, wm) = True.
\end{equation} 

In dataset watermarking, ownership verification can be done in either white- or black-box settings. In the white-box setting, a \emph{verifier} that can be \datasetowner or a judge (external, trusted party that runs the verification process) has access to the entire $\tilde{F}$. In the black-box setting, $\tilde{F}$ can only be accessed through a prediction API. 

\newtext{Radioactive data~\citep{sablayrolles2020radioactive} is a recent method to trace image classification datasets that are used to train DNN models back to their original owner. Although radioactive data is intended for data provenance and not for ownership verification, it can be also used in adversarial settings.} Radioactive data aims to find a minimally modified version $\tilde{x}_{wm}$ for each image $x_{wm} \in D_{wm}$ in order to intentionally transfer this knowledge to $\tilde{F}$ trained using the watermarked dataset $\tilde{D}$. $\tilde{x}_{wm}$ is generated using a marker DNN model $F_{m}$, and optimizing the objective function
\begin{align}
 \mathcal{L}(\tilde{x}_{wm}) = & -(\phi_m(\tilde{x}_{wm}) - \phi_m(x_{wm}))^{T}u \\
 & + \lambda_{1}\|  \tilde{x}_{wm} - x_{wm} \|_{2} + \lambda_{2}\| \phi_m(\tilde{x}_{wm}) - \phi_m(x_{wm})  \|_{2}, \nonumber
\end{align}\label{eq:radio}
\newtext{where $\lambda_{1}$ and $\lambda_2$ are constants penalizing $L_{2}$ distances over both pixel and feature space, $\phi_m$ is the feature extractor of $F_m$, and $u$ is the \emph{carrier} that shifts $\tilde{x}_{wm}$ in its direction. $u$ is a unit vector generated randomly, i.e. $\lVert u \rVert_{2} = 1$.}

In radioactive data~\cite{sablayrolles2020radioactive}, the detection, i.e., ownership verification in adversarial settings, is implemented in both white- and black-box settings. In white-box verification, first $\phi_m$ and and the feature extractor $\phi_s$ of the suspected DNN model $F_s$ are aligned, and then a hypothesis testing is applied by defining the following hypotheses: 
\begin{itemize}
    \item \textit{Null hypothesis H0: $\zeta_s$ was trained with the clean data.}%The carrier $u$ is independent of the weights of the linear classifier $\zeta$} 
    \item \textit{Alternative hypothesis H1: $\zeta_s$ was trained with the watermarked data.}%The carrier $u$ is independent of the weights of the linear classifier $\zeta$} 
\end{itemize}
%then a hypothesis testing is applied to check if the weights of the classifier $\mathcal{C}_s$ of $F_s$ is aligned with the carrier $\mathbf{u}$. 

\newtext{The cosine similarity between a fixed vector $v$ and another uniformly distributed vector $w$ on the unit sphere follows a beta-incomplete distribution~\cite{iscen2017memory}. Therefore, a statistical significance testing can be applied using the carrier $u$ and the hypotheses defined above.} If the value of cosine similarity between $u$ and weights of $\zeta_s$ is high, then the corresponding p-value (the probability of rejecting the null hypothesis when it is in fact true) is low, and there is enough evidence to refute \textit{H0}. Therefore, the verifier can claim that $\tilde{D}$ is used to train the suspected $F_s$ with a high confidence. For white-box verification, we chose the significance level as 5\%, such that if the p-value is smaller than $0.05$ (or $\log_{10}(p) \leq -1.3$), we reject \textit{H0} and verify the ownership of $\tilde{D}$.

In black box verification~\cite{sablayrolles2020radioactive}, the difference between the cross entropy loss $\ell$ of $x_{wm} \in D_{wm}$ and $\tilde{x}_{wm} \in \tilde{D}_{wm}$ is calculated as 

\begin{equation}\label{eq:blackboxeqn}
    1/|D_{wm}|\sum_{i=0}^{|D_{wm}|}\ell(x_{wm_i}, c_{i}) - \ell(\tilde{x}_{wm_i}, c_{i}),
\end{equation}
where $c_{i}$ is the true class for the i-th sample. % The cross entropy loss is calculated using the output prediction vector $F_s$ and a one-hot version of the true class $c$. 
If the output of Equation~\ref{eq:blackboxeqn} is higher than zero, it shows that $F_s$ fits better to $\tilde{D}_{wm}$ than $D_{wm}$, and the verifier can demonstrate the ownership of $\tilde{D}$. % average cross entropy loss for radioactive samples is lower than their clean version, $F_s$ was possibly trained with $D_v$.

\subsection{Model Watermarking}\label{ssec:modelwatermarking}

Model watermarking is a well-known strategy to trace unauthorized, stolen copies back to the original owner. Model watermarking consists of two phases: \emph{embedding} and \emph{verification}. In the embedding phase, the watermark is either inserted into the model weights directly, or learned through a watermark set (i.e., watermarking via backdooring~\cite{adi2018turning}). In the verification phase, the model owner needs to satisfy predefined requirements to successfully demonstrate the ownership. These requirements could include scoring a low bit error rate (BER) between the suspected and non-watermarked model weights~\cite{darvish2019deepsigns, chen2019blackmarks,chen2020specmark}, or achieving a watermark accuracy higher than some threshold value~\cite{adi2018turning, zhang2018protecting, li2019prove}.

\subsection{Model Extraction Attacks and Ownership Demonstration of ML Models}\label{ssec:modelextraction}

In recent years, instead of 
distributing ML models to users, model owners host their models in centralized servers and allow users to query the model via a prediction API. Although prediction APIs protect the direct leakage of ML models, adversaries can mount model extraction attacks to steal the functionality of ML models using only prediction vectors~\cite{tramer2016stealing,orekondy2019knockoff,correia2018copycat,atli2020extraction}. In model extraction attacks, the adversary $\mathcal{A}$ first obtains a set of unlabeled data $D_{\mathcal{A}}$, which contains either natural or synthetically generated samples. Then, $\mathcal{A}$ queries the API of a victim model $F_{\mathcal{V}}$ to obtain a set of pseudo-labels $F_{\mathcal{V}}(D_{\mathcal{A}})$ for $D_{\mathcal{A}}$. $\mathcal{A}$ uses the transfer set $\{D_{\mathcal{A}}, F_{\mathcal{V}}(D_{\mathcal{A}})\}$ to train a surrogate model $F_{\mathcal{A}}$ which performs relatively similar to the victim model $F_{\mathcal{V}}$. Although some detection~\cite{atli2020extraction} and prevention methods~\cite{orekondy2019prediction} are effective in different adversary models, they either negatively affect the model performance and the utility of benign users~\cite{atli2020extraction}, or fail against $\mathcal{A}$ with realistic assumptions: 1) $\mathcal{A}$ knows the task of $F_{\mathcal{V}}$, 2) collects a large number of natural dataset $D_{\mathcal{A}}$ related to the task, 3) has access to pre-trained ImageNet models, and 4) uses the most confident label(s) to train $F_{\mathcal{A}}$.
 
Instead of preventing model extraction attacks, different ownership verification methods including model watermarking are proposed to trace back the ownership of the surrogate model $F_{\mathcal{A}}$ to the original model $F_{\mathcal{V}}$. However, many model watermarking techniques are found to be not robust against adaptive attacks~\cite{chen2021refit, wang2019neural, shafieinejad2021robustness} or model extraction attacks~\cite{chen2019leveraging, lukas2022watermarkingsok} that generate $F_{\mathcal{A}}$ with modified decision boundaries. Recently, alternative ownership verification methods such as dynamic watermarking~\cite{szyller2019dawn}, fingerprinting~\cite{zhao2020afa, lukas2019deep} and dataset inference~\cite{maini2021dataset} have been proposed to overcome these issues. Instead of embedding the watermark into ML models, these methods use the knowledge transferred by the model or the original training data. However, these methods also have their own limitations considering different security and privacy requirements. For example, fingerprinting methods~\cite{zhao2020afa, lukas2019deep} are more robust than model watermarking, but they require generating or training multiple reference models to extract transferable adversarial examples that are used as fingerprints. Similarly, dataset inference does not need an additional phase for model training or fingerprint generation, but it is at odds with privacy since it leaks information about the training data during the verification process.
\section{Problem Setting and Goals}\label{sec:adversarymodel}
In this section, we define an adversary model and list the requirements for an effective ownership demonstration for any dataset watermarking technique. %Then, we present our reproduction results and investigate the effectiveness of radioactive data using different datasets. 

\noindent\textbf{Adversary model}: The adversary model of dataset watermarking is different from traditional model watermarking methods, since the knowledge available to owners and adversaries are different in these setups. In model watermarking, the model owner has the original dataset $D$, generates its own watermark $wm$, and trains a watermarked model $\tilde{F}$. %so that a verification algorithm \textsc{Verify} will result in \emph{True}. 
The extent of $\mathcal{A}$'s knowledge of $D$ depends on the setting and can range from full access to $D$~\citep{adi2018turning}, through partial access (to a subset of $D$)~\citep{wang2019neural,aiken2021neural}, to no access beyond knowledge of $D$'s domain~\citep{chen2019leveraging}. In dataset watermarking, the original dataset owner \datasetowner has $D$, and generates $wm$ and the watermarked dataset $\tilde{D}$, but $\mathcal{A}$ can select any algorithm and model architecture to train a model $\tilde{F}_{\mathcal{A}}$ using $\tilde{D}$, and has a complete knowledge of $\tilde{D}$. $\mathcal{A}$ 's goal is to train $\tilde{F}_{\mathcal{A}}$ using $\tilde{D}$ such that (1) $F_{\mathcal{A}}$ approximates the oracle $\mathcal{O}_f$, and (2) monetize $F_{\mathcal{A}}$ in an unauthorized way. \datasetowner's objective is to verify the ownership of $\tilde{D}$ that is used to build $\tilde{F}_{\mathcal{A}}$ and prove that there is a misuse. %(2) evades the ownership verification of $D_v$ using $(v,w)$.

%\noindent\textbf{Adversary's Capabilities}: $\mathcal{A}$ has full control over $\tilde{D}$. However, 

Watermarked samples $\tilde{x}_{wm} \in \tilde{D}_{wm}$, clean version of these samples $x_{wm} \in D_{wm}$, and $wm$ are only known by \datasetowner. $\mathcal{A}$ can apply data pre-processing techniques in order to remove the effect of $wm$. Second, $\mathcal{A}$ controls the training and can modify the training algorithm in order to prevent the embedding of $wm$ in $\tilde{F}_\mathcal{A}$. The training hyper-parameters (e.g., batch size, learning rate) can be set at will and regularization techniques (e.g., weight decay, drop out, etc.) can combined during training. Finally, $\mathcal{A}$ chooses any architecture it wishes as its model. %$\mathcal{H}$ is restricted by the victim to a type of model architecture for which transitive watermarking properties are guaranteed. 
%$\mathcal{H}$ can be more or less specific, e.g., ResNet DNN model family~\citep{he2016deep}, $\rightarrow$ ImageNet models $\rightarrow$ Convolutional Neural Networks (CNN) $\rightarrow$ Deep Neural Networks (DNN). 

\noindent\textbf{Assumptions:} Similar to the setup in radioactive data~\cite{sablayrolles2020radioactive}, we assume that the complete $\tilde{D}$ is used for the training and is not mixed with other training sets. In addition, due to different verification mechanisms explained in~\ref{ssec:datasetwm}, we assume that $\mathcal{A}$ monetizes $\tilde{F}_{\mathcal{A}}$ by making it available either in plain-text or via its prediction interface to its own customers. Anyone who wants to verify the ownership of the dataset used to train $\tilde{F}_{\mathcal{A}}$ can do so by becoming a customer of $\mathcal{A}$ thereby gaining access to $\tilde{F}_{\mathcal{A}}$ itself (white-box verification) or its prediction interface (black-box verification). 
%$\mathcal{A}$ publicly deploys $\tilde{F}_{\mathcal{A}}$ in a prediction API, or can make $\tilde{F}_{\mathcal{A}}$ available to parties that are responsible for implementing white-box verification. 

\noindent\textbf{Requirements:} Based on the adversary model and the assumptions explained above and inspired by the prior work~\cite{uchida2017embedding,adi2018turning, lukas2022watermarkingsok}, we define the following requirements for a successful %designing an effective 
ownership verification:
\begin{enumerate}[leftmargin=*]
    \item \textbf{Utility:} The maximum acceptable test accuracy drop between $F$ ($F$ trained using the original, unmodified $D$) and $\tilde{F}$ should be lower than $5pp$, i.e., $ Acc(F) - Acc(\tilde{F}) \leq 5pp$.
    \item \textbf{Effectiveness:} \newtext{Irrespective of the model architecture and training algorithm, $\textsc{Verify}(\tilde{F}, D_{wm}, wm)$ should return $True$ for any $\tilde{F}$ trained with $\tilde{D}$.}% For any $\tilde{D}$, \datasetowner should be able to identify its dataset that used to train ML models with high confidence. Irrespective of the model architecture and training algorithm, $\textsc{Verify}(\tilde{F}, D_{wm}, wm)$ should return $True$ for any $\tilde{F}$ trained with $\tilde{D}$.
	%\item \textbf{Generality:} Irrespective of the model architecture and training algorithm, $\textsc{Verify}(\tilde{F}, D_{wm}, wm)$ should return $True$ for any $\tilde{F}$ trained with $\tilde{D}$.
	\item \textbf{Integrity:} \datasetowner should avoid wrongly accusing any model $F$ that is trained with other datasets $D^{'}$, including $D$, i.e. $\textsc{Verify}$ should return $False$. %Additionally, if $F$ is trained with the original dataset $D$, the model owner should not be able to demonstrate the ownership for $F$, %i.e., $\textsc{Verify}({F}, D_{wm}, wm)$ should return $False$.
	\item \textbf{Stealthiness:} Watermarked samples $\tilde{x}_{wm} \in \tilde{D}_{wm}$ should be visually similar to their clean versions $x_{wm}$ such that $\mathcal{A}$ cannot detect any $\tilde{x}_{wm}$ through a visual inspection. In addition, watermarked samples should not be detected easily via simple clustering or out-of-distribution detectors. 
	\item \textbf{Robustness}: Watermarked samples should be robust against input transformations such as rotation and re-scaling. 
\end{enumerate}
%We assume the adversary publicly deploys $F$ in a prediction API. A defender can query this model to obtain predictions from it. This assumption is common to any watermarking solution where the watermarked media must be accessible by the owner to prove the watermark is embedded in it. This is reasonable since an illegally acquired media harms its creator's rights the most when it is widely distributed. 

In backdoor-based dataset watermarking~\cite{li2020open}, each $\tilde{x}_{wm}$ is reassigned with a different label. Since these samples are far away from their true classes, these samples can be detected by simple clustering methods, thus failing to satisfy the stealthiness requirement. Moreover, any dataset ownership verification method that relies on backdoor-based watermarking suffers from the same limitations (i.e., recovering the watermark pattern~\cite{wang2019neural}, reverse engineering watermarked samples~\cite{wang2019neural,aiken2021neural}) as in backdoor-based model watermarking. This confirms that backdoor-based watermarking is not an effective method for demonstrating the ownership of datasets.

%auto-ignore
\section{Radioactive Data: Evaluation}\label{sec:evaluation}

\newtext{In this section, we evaluate the effectiveness of radioactive data~\cite{sablayrolles2020radioactive}, which is the first data provenance method that can be used for dataset watermarking.} 

\subsection{Experimental Setup}\label{ssec:expsetup}
To evaluate the effectiveness of radioactive data~\cite{sablayrolles2020radioactive}, we begin with the CIFAR10 and CIFAR100 datasets~\citep{krizhevsky2009learning}. %CIFAR10~\citep{krizhevsky2009learning} contains 60,000 RGB images divided into 10 different classes with 5000 images per class, and 10,000 images are reserved for the test set. CIFAR100~\citep{krizhevsky2009learning} also contains 60,000 RGB images, 100 classes (500 per class), and 10,000 images are reserved for the test set. 
We also constructed CIFAR10$^{*}$, CIFAR30, and CIFAR50 by collecting samples belonging to 10, 30, and 50 classes, respectively, from CIFAR100.

We used ResNet-18~\citep{he2016deep} for the marker model $F_m$, and all models $\tilde{F}_{\mathcal{A}}$ trained with the watermarked dataset $\tilde{D}$, since ResNet-18 was also used in the original paper~\cite{sablayrolles2020radioactive}. %In addition to the contaminated models with the ResNet-18 architecture, 
We also trained reference models $F_r$ on the original CIFAR10 dataset $D$ to evaluate the integrity requirement. We chose ResNet-18, DenseNet-121~\citep{huang2017densely} and AlexNet~\citep{krizhevsky2009learning} for training $F_r$. For the feature extractor $\phi(x)$, we used the penultimate layer of $F$ just before the classification layer, similar to the original paper~\citep{sablayrolles2020radioactive}.

We downloaded the GitHub repository for radioactive data\footnote{\url{https://github.com/facebookresearch/radioactive_data}}, and modified the code to extend dataset watermarking experiments for CIFAR10, CIFAR100 and different subsets of CIFAR100. For generating $x_{wm}$, we used the default parameters suggested in the public repository. %Before training $F_{m}$ and $\tilde{F}_{\mathcal{A}}$, all images were resized to $3 \times 256 \times 256$, cropped to $3 \times 224 \times 224$, and finally normalized with mean and standard deviation values suggested for these datasets~\citep{krizhevsky2009learning}. 
We trained $F_{m}$ and $\tilde{F}_{\mathcal{A}}$ \newtext{using the same setup in the GitHub repository. }%for 120 epochs with the batch size of 64 using stochastic gradient descent (SGD) with a learning rate of $0.1$, momentum of $0.9$ and weight decay of $0.0001$. We also used a decaying learning rate to decrease the learning rate by 10 in every 30 epochs. %In backdoor-based watermarking, we trained $F_r$, $F_v$ and $F_{\mathcal{A}}$ for 200 epochs with a batch size of 50. Similar to radioactive watermarking, we also used SGD in optimization procedure with a fixed learning rate of 0.1, momentum of 0.9, and weight decay of 0.0005. 
We experimented with different watermarking ratios $wm_r$ (the proportion of the watermarked samples relative to total number of samples in one class) in order to identify possible effects of it on $\tilde{F}_{\mathcal{A}}$. %We performed our experiments using Python 3.6 with Numpy 1.17, PyTorch 1.4, torchvision 0.2.1. All experiments are implemented on a computer with NVIDIA Tesla V100 GPU.  

%We downloaded GitHub repositories for radioactive\footnote{\url{https://github.com/facebookresearch/radioactive_data}} and backdoor-based watermarking\footnote{\url{https://github.com/THUYimingLi/Open-sourced_Dataset_Protection}} techniques, and modified them to extend dataset watermarking experiments for CIFAR10 and CIFAR100. In backdoor-based watermarking, we wrote additional data loader functions for CIFAR100. We also used the original ResNet-18 architecture~\citep{he2016deep} instead of using the modified, smaller ResNet-18, which was the default model architecture in the GitHub repository. In radioactive-watermarking, we modified the code so that it could easily load different datasets and obtain watermarked samples using the image paths. For watermarking data samples, we used the default parameters suggested in these public repositories. %The corresponding public repositories are : \url{https://github.com/facebookresearch/radioactive_data} for radioactive watermarking, and \url{https://github.com/THUYimingLi/Open-sourced_Dataset_Protection} for backdoor-based watermarking. 
%We performed our experiments using the same setup as in the originals papers. Setup 1 (Python 3.6 with Numpy 1.17, PyTorch 1.4, torchvision 0.2.1, NVIDIA Volta V100 GPU with 32GB of memory) was used for radioactive watermarking, and Setup 2 (Python 3.8 with Numpy 1.20, PyTorch 1.8, torchvision 0.9.1, NVIDIA A100 GPU with 40GB VRAM) was used for implementing backdoor-based watermarking. We should note that \ourmethod is optimized to work in both setups.

\subsection{Reproduction Results}\label{ssec:reproduction}

In the original paper~\cite{sablayrolles2020radioactive}, radioactive data was shown to be robust against input transformations. Therefore, we focus on evaluating the four remaining requirements using different image datasets. Table~\ref{table:radioactivedata} summarizes our experimental results. 

\begin{table*}[t]
	\centering
	\resizebox{1.0\textwidth}{!}{%
		\begin{tabular}{c c c c c c } 
			\hline
			  %&  watermarking ratio &  test accuracy &  white-box  & black-box & white-box \\ 
			 Dataset & watermarking ratio $wm_r$ &  test accuracy $Acc(\cdot)$ & white-box verif. w/ $D_{test}$ & black-box ver. & white-box verif. w/ $\tilde{D}_{wm}$ \\
			\hline
			 & marker & 87.27\%  & \color{green} -0.480 & \color{green} -0.275 & \color{green} -0.480 \\
			 
			 CIFAR10 (5000 images & 10\% & 86.81\% & \color{green} -2.804 & \color{green} 0.171  & \color{green} -9.563 \\
			 
		       per class) & 20\% & 85.95\% & \color{green} -1.835 & \color{green} 0.260  &  \color{green} -12.098 \\
		      \hline
		      
		      & marker & 85.17\% & \color{green} -0.508 & \color{green} -3.430  & \color{green} -0.508 \\
		      
		      CIFAR10$^*$ (500 images  & 10\% & 86.97\%  & \color{red} -0.484 & \color{green} 0.022 & \color{red} -0.386\\
		      
		       per class) & 20\% & 86.03\% & \color{red} -0.249 & \color{green} 0.023 & \color{red} -0.863 \\
		      \hline
		      
		      & marker & 76.70\% & \color{green} -0.361 & \color{green} -0.667 & \color{green} -0.361\\
		      
		      CIFAR30 (500 images & 10\% & 76.51\% & \color{red} -0.411 & \color{green} 0.048 & \color{green} -3.214 \\
		  
		      per class) & 20\% & 73.40\% & \color{red} -0.266 & \color{green} 0.057 & \color{green} -9.177 \\
		      \hline
		      
		      & marker & 69.83\% & \color{green} -0.396 & \color{green} -0.992  & \color{green} -0.396 \\
		      
		      CIFAR50 (500 images & 10\% & 65.64\% & \color{green} -1.614 & \color{green} 0.077  & \color{green} -21.317 \\
		      
		       per class) & 20\% & 65.76\% & \color{green} -5.779 & \color{green} 0.172 & \color{green} -26.183 \\
		      \hline
		      
		      & marker & 61.84\% & \color{green} -0.176 & \color{green} -2.098 &  \color{green} -0.176\\
		      
		      CIFAR100 (500 images & 10\% & 61.62\%  & \color{green} -4.894 & \color{green} 0.277 & \color{green} -72.113 \\
		      
		       per class) & 20\% & 60.82\%  & \color{green} -9.556 & \color{green} 0.467 & \color{green} -102.160 \\
			\hline
	\end{tabular}}
	\caption{White- and black-box ownership verification results of radioactive data for clean models $F_m$ used to generate $D_{wm}$, and $\tilde{F}_{\mathcal{A}}$ trained using $\tilde{D}$. Expected and unexpected verification results are highlighted in green and red, respectively.}\label{table:radioactivedata}
\end{table*}

\textbf{Utility:} \newtext{Column 3 (``test accuracy $Acc(\cdot)$'') in Table~\ref{table:radioactivedata} presents that the maximum difference of test accuracy between $F_{m}$ and $\tilde{F}_{\mathcal{A}}$ is below $5$pp. Therefore, radioactive data satisfies the utility requirement.} %We also observed that if the watermarking ratio increases from 10\% to 20\%, the test accuracy decreases. Therefore, datasets owners should decide the best watermarking ratio considering the utility constraint.

\textbf{Effectiveness:}
Columns 4-6 in Table~\ref{table:radioactivedata} show the effectiveness of radioactive data in different settings. Column 4 (``white-box verification with $D_{test}$'') and Column 5 (``black-box verification'') correspond to our reproduction of white-box and black-box verification techniques (respectively) from~\cite{sablayrolles2020radioactive} evaluated in our experimental setup. Column 4 presents the $\log_{10}(p)$ value for white-box verification implemented with the test data $D_{test}$, as suggested in the original paper~\cite{sablayrolles2020radioactive}. %, and column 5  shows the black-box verification. 
The last column is discussed in Section~\ref{ssec:improvedwhitebox}.

Column 4 shows that white-box verification~\cite{sablayrolles2020radioactive} is not effective in all settings, although the verifier has a complete access to $\tilde{F}_{\mathcal{A}}$. We implemented white-box verification using the test dataset $D_{test}$, since the original paper~\cite{sablayrolles2020radioactive} also uses the validation set of ImageNet~\cite{deng2009imagenet}. Column 4 shows that for a fixed number of classes (CIFAR10 vs CIFAR10$^{*}$), if the sample size per class is small ($\leq 500$), there is not enough evidence to reject \textit{H0} when the significance level is 5\% (i.e., $\log_{10}(p) > -1.3$), and the ownership verification fails. Similarly, for a fixed number of samples per class, if the number of classes is low $\leq 30$, then white-box verification fails. 
%Table~\ref{table:radioactivedata} presents the complete white- and black-box verification results. 

In contrast to white-box verification, black-box verification~\cite{sablayrolles2020radioactive} results (Column 5) confirm that for any $\tilde{D}$, and watermarking ratio $wm_r$, the loss difference formalized in Section~\ref{ssec:datasetwm} is always bigger than zero. Therefore, black-box verification~\cite{sablayrolles2020radioactive} is always effective. We also observed that if the dataset has a high number of classes or a bigger $wm_r$, the loss difference increases, thus leading to better confidence during the ownership verification.

\textbf{Integrity:} To check the integrity requirement, we apply white- and black-box verification on the marker model $F_{m}$ that contains no watermarked samples. Columns 4 (white box verification with $D_{test}$~\cite{sablayrolles2020radioactive}) and 5 (black-box verification~\cite{sablayrolles2020radioactive}) in Table~\ref{table:radioactivedata} show that we cannot verify the ownership of $F_{m}$, as expected. Furthermore, Table~\ref{table:referencemodels} shows that despite the different model architecture, \datasetowner avoids falsely accusing models trained using $D$ without any watermark. Radioactive data satisfies the integrity requirement.

\begin{table}[t]
	\centering
	\resizebox{1.0\linewidth}{!}{%
		\begin{tabular}{ l l c c c } 
			\hline
			 Reference & test accuracy & white-box & black-box & white-box\\ 
			 models $F_{r}$ & $Acc(\cdot)$  & verif. w/$D_{test}$ & verification & verif. w/$D_{wm}$\\
			\hline
			ResNet-18   &  87.54\% &  -0.284 & -2.906 & -0.337\\ %-3.059 -- -2.906\\ 
			AlexNet     &  85.88\%&  -0.272 & -0.168 & -0.266 \\ %-2.666 -- - 0.168\\ 
			DenseNet-121 & 90.99\% &  -0.910 & -2.570 & -0.842 \\ %-2.628 -- -2.570\\ 
			\hline
	\end{tabular}}
	\caption{White- and black-box ownership verification results of radioactive data for clean models $F_{r}$ trained with the original, non-watermarked CIFAR10 dataset, where the watermarking ratio $wm_{r}$ is set to 10\% and 20\%.}
	\label{table:referencemodels}
\end{table}

\begin{figure}[t]
	\centering
	\subfloat[ \label{fig:s1}]{\includegraphics[width=0.20\columnwidth]{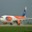}}
	\hspace{0.3cm}
	\subfloat[\label{fig:s2}]{\includegraphics[width=0.20\columnwidth]{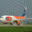}}
	\hspace{0.3 cm}
	%\;
	\subfloat[\label{fig:s5}]{\includegraphics[width=0.20\columnwidth]{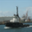}}
	\hspace{0.3 cm}
	\subfloat[\label{fig:s6}]{\includegraphics[width=0.20\columnwidth]{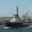}}
	\caption{Visualization of clean and watermarked images for CIFAR10 samples. We show the original images (a) for class ``airplane'' and (c) for class ``ship''; watermarked images via radioactive watermarking (b) and (d) respectively.} \label{fig:samplesets}
\end{figure}

\textbf{Stealthiness:} \newtext{The original paper~\cite{sablayrolles2020radioactive} proves that for ImageNet samples, the visual distortion is undetectable. We also compared clean and watermarked images from CIFAR10 in Figure~\ref{fig:samplesets}. %, since images in CIFAR10 has smaller image size than ImageNet.
%Figure~\ref{fig:samplesets} shows that $wm$ is different for each radioactive image. 
Although some watermarked samples can be recognized with visual inspection, additive watermarks resemble random noise, so it might be difficult to identify every $x_{wm}$.} %In images with bigger dimensions, it would be more difficult to detect watermarked images, as shown in the original paper~\citep{sablayrolles2020radioactive}. 

\subsection{Improved White-box Verification}\label{ssec:improvedwhitebox}
With white-box verification, a verifier will have strictly greater access to a suspect model than with black-box verification. Therefore, the results we observed in Section~\ref{ssec:reproduction}, showing that the black-box verification succeeds even in settings where white-box verification fails, are counter-intuitive. \newtext{This suggests that the white-box technique described in~\cite{sablayrolles2020radioactive} is not an optimal ownership verification technique considering adversarial settings.}

Based on the hypothesis testing explained in Section~\ref{ssec:datasetwm}, we conjecture that directly using $\tilde{D}_{wm}$ instead of $D_{test}$ can decrease the p-value and increase the confidence in white-box verification. \newtext{Since the cosine similarity between $u$ and weights of $\zeta_{\mathcal{A}}$, which is aligned with $\phi_{\mathcal{A}}(\tilde{x}_{wm})$, should be high when $\tilde{D}_{wm}$ is used in the verification instead of $D_{test}$. Then, the corresponding p-value will be lower and there will be enough evidence to refute the null hypothesis. }%Since each $\tilde{x}_{wm} \in \tilde{D}_{wm}$ is a subset of the training set $\tilde{D}$ and shifted by the direction $u$ (Equation~\ref{eq:tw_dataset}), the cosine similarity between $u$ and weights of the linear classifier of adversary's DNN model $\zeta_{\mathcal{A}}$ should be higher than $x \in D_{test}$. 
The last column in Table~\ref{table:radioactivedata} (``white-box verif. w/ $\tilde{D}_{wm}$'') confirms our conjecture. Compared to $D_{test}$, white-box verification with $\tilde{D}_{wm}$ gives lower p-values and increases the confidence of the verification. Updated results presents that white-box verification only fails in CIFAR10$^{*}$ case at a 5\% significance level. \newtext{There is a positive correlation between the effectiveness of radioactive data and the dataset size, and radioactive data may not be a suitable ownership demonstration method for small datasets.} %This shows that in radioactive data, if the number of classes is low, the size of the watermarked set should be high enough for later correctly verifying the ownership.

\newtext{Since there is no statistical guarantee for the black-box verification, we suggest a proper use of the ownership verification as follows: Even if the verifier has the white-box information, it should first run black-box verification on suspected models due to its higher effectiveness. Then, if the verifier gains a white-box access, it should report $\log_{10}(p)$ obtained by white-box verification using $\tilde{D}_{wm}$ to further support the decision with a confidence value or double-check its judgement when black-box verification fails.}
%auto-ignore
\section{Radioactive Data vs. Model Extraction Attacks}\label{sec:againstmodelextraction}

In this section, we explore the effectiveness of radioactive data in the presence of model extraction attacks.
\begin{table*}[t]
	\centering
	\resizebox{1.0\textwidth}{!}{%
		\begin{tabular}{c c c c c c c} 
			\hline
			  %& $wm_r$ &  test accuracy & $ Acc(\tilde{F}_\mathcal{A})  -$ &white-box  & black-box & white-box \\ 
			  Dataset & $wm_r$ of $\tilde{F}_{\mathcal{A}}$  & Test accuracy $Acc(.)$ 
			  & $Acc(\tilde{F}_\mathcal{A}) - Acc(F^{*}_{\mathcal{A}}) $ & white-box ver. verif. w/ $D_{test}$ & black-box verif. & white-box verif. w/ $D_{wm}$ \\
			  \hline
			  
			  CIFAR10 (5000 images & 10\% & 82.38\% & 4.43 pp & \color{green} -1.537 & \color{green} 0.160 &  \color{green} -4.042\\
		       per class ) & 20\% & 80.34\% & 5.61 pp & \color{green} -2.327 & \color{green} 0.240  & \color{green}-3.256 \\
		      \hline
		      
		      CIFAR10$^*$(500 images  & 10\% & 85.67\% & 1.3 pp & \color{red} -0.150 & \color{green} 0.034 & \color{red} -0.561\\
		      per class)  & 20\% & 86.05\% & -0.1 pp  & \color{red} -0.132 & \color{green} 0.062 & \color{red} -1.013\\
		      \hline
		      
		      CIFAR30 (500 images & 10\% & 75.44\% & 1.07 pp & \color{red} -0.259 & \color{green} 0.002 & \color{green} -1.453 \\
		      per class) & 20\% & 72.01\% & 1.39 pp & \color{red} -0.908 &  \color{green} 0.071 &  \color{green} -1.490 \\
		      \hline
		      
		      CIFAR50 (500 images & 10\% & 59.17\%  & 4.72 pp & \color{red} -1.185 & \color{red} -0.020 &  \color{green} -1.756 \\
		      per class)  & 20\% & 63.92\% & 1.84 pp & \color{green} -3.345 & \color{green} 0.143 & \color{green} -3.819 \\
		      \hline
		      
		      CIFAR100 (500 images & 10\% & 54.76\% & 6.86 pp  & \color{green} -2.622 & \color{red} -0.033 & \color{green} -8.276 \\
		      
		      per class)  & 20\% & 53.93\% & 6.89 pp  & \color{green} -4.364 &\color{green} 0.198 & \color{green} -19.274 \\
	\end{tabular}}
	\caption{White- and black-box ownership verification results for surrogate models $F^{*}_{\mathcal{A}}$.  $F^{*}_{\mathcal{A}}$'s are obtained by implementing a Knockoff attack on adversary's models $\tilde{F}_{\mathcal{A}}$ trained via $\tilde{D}$. Successful and failed verifications are highlighted in green and red, respectively.}\label{table:modelstealing}
\end{table*}

\subsection{Can Model Extraction Thwart Dataset Watermarking with Radioactive Data?}\label{ssec:radioactive-modelextraction}
%\noindent{\textbf{Radioactive data watermarks are resistant against watermark removal via model extraction attacks.}}
Fine-tuning~\cite{szyller2019dawn, lukas2022watermarkingsok, chen2019leveraging} is one of the most common watermark removal methods that is implemented against model watermarking strategies. Similarly, in dataset watermarking, an adversary $\mathcal{A}$ can mount a model extraction attack against its own model $\tilde{F}_{\mathcal{A}}$ to obtain a ``surrogate'' model $F^{*}_{\mathcal{A}}$ in an attempt to remove any watermarks to evade the ownership verification of the dataset $\tilde{D}$. 

To evaluate this, we use Knockoff nets~\cite{orekondy2019knockoff} a state-of-the-art model extraction attack on complex DNNs. In Knockoff nets, $\mathcal{A}$ queries $\tilde{F}_{\mathcal{A}}$ using a transfer set $D_{\mathcal{A}}$, which is not related to the original task but composed of natural samples collected from online databases, and minimizes the KL divergence between $F^{*}_{\mathcal{A}}$ and $\tilde{F}_{\mathcal{A}}$'s predictions. As $\tilde{F}_{\mathcal{A}}$, we used models presented in Table~\ref{table:radioactivedata}. Similar to the original experimental setup in Knockoff nets~\cite{orekondy2019knockoff}, for constructing $D_{\mathcal{A}}$, we sampled 100,000 images from the ImageNet dataset, 100 images per class. We trained surrogate models $F^{*}_{\mathcal{A}}$ using SGD optimization, with an initial learning rate of 0.01 that is decreased to 0.001 after 60 epochs, and an overall 100 epochs. 

Table~\ref{table:modelstealing} summarizes white- and black-box verification values for $F^{*}_{\mathcal{A}}$. Column 5 (``white-box ver. w/$D_{test}$'') shows that white-box verification using $D_{test}$ is not effective for every setting, while column 7 (``white-box ver. w/$D_{wm}$'') demonstrates that using $D_{wm}$ verifies the ownership of all $F^{*}_{\mathcal{A}}$'s, except CIFAR10$^{*}$. The exception in CIFAR10$^{*}$ is intuitive, since the verification also fails for $\tilde{F}_{\mathcal{A}}$ trained via watermarked  CIFAR10$^{*}$ (Column 6 in Table~\ref{table:radioactivedata}). The black-box verification also succeeds in many cases, except CIFAR50 and CIFAR100 with a watermarking ratio of 10\%. When these two cases were investigated further, it was observed that $F^{*}_{\mathcal{A}}$'s do not recover the performance of $\tilde{F}_{\mathcal{A}}$'s, and achieve a low test accuracy. Therefore, we conclude that if there is no significant decrease in the test accuracy, then radioactive data watermarks are retained even after an attempt to remove those via model extraction.

\subsection{Can Radioactive Data be Used for Model Watermarking?}\label{ssec:radioactive-modelwatermarking}

%In model extraction attacks~\cite{tramer2016stealing,orekondy2019knockoff,correia2018copycat}, the adversary $\mathcal{A}$ queries victim model's $F_{\mathcal{V}}$ prediction API, and exploits responses in order to build a surrogate model $F_{\mathcal{A}}$ with a similar performance of $F_{\mathcal{V}}$. 
Recent work~\cite{chen2019leveraging, lukas2022watermarkingsok} shows that existing model watermarking methods cannot preserve the watermarks in model extraction attacks, since these attacks modify decision boundaries. Therefore, a different ownership verification strategy should be designed to trace the surrogate model back to the original model. Maini \textit{et  al.}~\cite{maini2021dataset} shows that $F_{\mathcal{A}}$ contains direct or indirect knowledge from the victim's $\mathcal{V}$ training set, \newtext{and they use this observation for proposing \emph{dataset inference}, a model ownership verification method that identifies the knowledge in the training set transferred from original to the stolen model.} Based on this observation and our findings in Section~\ref{ssec:radioactive-modelextraction}, we suggest that radioactive watermarking can serve as an alternative ownership verification technique resistant to model extraction attacks, providing the ability to trace surrogate models obtained via model extraction back to the original model.

\begin{figure}[t]
    %\centering
    %\subfloat[white-box ver. w/ $D_{wm}$ ]{{\includegraphics[width=0.8\columnwidth]{figures/white_box_cifar.pdf} }}%
    %\quad
    \subfloat[ black-box verification]{{\includegraphics[width=0.8\columnwidth]{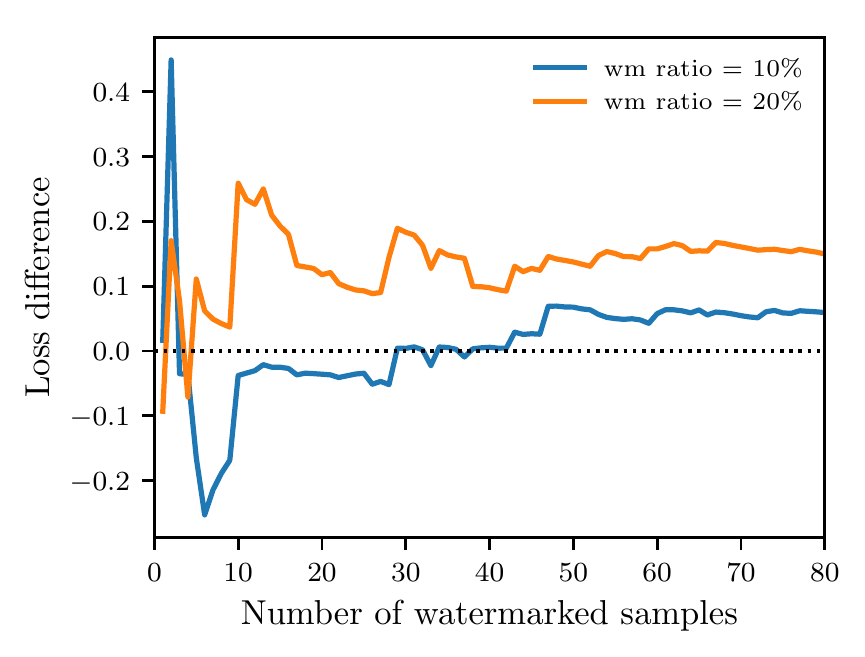} }}%
    \caption{%$-\log_{10}(p)$ value and loss difference used for white- and black-box verification against the number of watermarked samples used during the verification of CIFAR10 models. Revealing less than 50 watermarked samples is enough for the black-box verification.
    Loss difference used black-box verification against the number of watermarked samples used during the verification of CIFAR10 models. Revealing less than 50 watermarked samples is enough for the verification.}%
    \label{fig:example}%
\end{figure}

In this setup, we assume that the dataset owner $DO$ is the same as the model owner $\mathcal{V}$ or gives the right to $\mathcal{V}$ to monetize its model. The victim model $\tilde{F}_{\mathcal{V}}$ is trained with the watermarked dataset $\tilde{D}$ obtained using radioactive data. $\tilde{F}_{\mathcal{V}}$ is deployed on a centralized server and its prediction API returns the probability vector to each query. Figure~\ref{fig:example} shows the number of watermarked samples during the black-box verification process of $F^{*}_{\mathcal{A}}$ in CIFAR10. As shown in the figure, %white-box verification requires revealing many watermarked samples, while 
black-box verification has to reveal 100 samples (50 watermarked + 50 clean) where the watermarking ratio $wm_r$ is 10\%. If $wm_r$ is increased to 20\%, less than 20 samples (10 watermarked + 10 clean) is enough for a successful verification of ownership. \newtext{In addition, $\mathcal{A}$ cannot reverse engineer watermarks easily, since each embedded watermark $wm$ is different for every watermarked sample $\tilde{x}_{wm}$, and $\tilde{x}_{wm}$'s are not assigned to a different label. %Thus, radioactive data could be used a a query-efficient model ownership verification strategy. 
Thus, in black-box verification settings, radioactive data can function by revealing only a very small percentage of watermarked samples.}

%auto-ignore
\section{Conclusion}\label{sec:conclusion}
We reproduced and systematically evaluated the dataset provenance method in~\cite{sablayrolles2020radioactive}, radioactive data, in a variety of experiment settings. \newtext{We showed that while black-box verification from~\cite{sablayrolles2020radioactive} is effective in all settings, the same cannot be said about the white-box verification technique from~\cite{sablayrolles2020radioactive} but it can be improved.} We also showed that radioactive data survives model extraction attacks. This suggests that beyond dataset watermarking, radioactive data may potentially serve as a \emph{model watermarking} technique, resistant to model extraction. \newtext{In future work, we aim to further measure the robustness of radioactive data when it is used as an ownership verification method for both ML models and datasets by adding more quantitative analysis on variety of datasets and implementing state-of-the-art watermark removal attacks including fine-tuning~\cite{uchida2017embedding, chen2019leveraging}, fine-pruning~\cite{liu2018fine} and adversarial training~\cite{madry2017towards}.}

In our reproduction experiments, we used the white-box and black-box verification methods as in the original paper~\cite{sablayrolles2020radioactive}. %The $\textsc{Verify}$ process of he 
Black-box verification in~\cite{sablayrolles2020radioactive} relies on the verifier sending a mix of watermarked and clean samples to the suspected model. % Querying the suspected model with watermarked samples is the pattern used by $\textsc{Verify}$ in model watermarking. However, in dataset watermarking, %unlike in model watermarking, 
In dataset watermarking, the adversary $\mathcal{A}$ \emph{knows} the entire training set $\tilde{D}$ which includes watermarked samples. This may provide a way for $\mathcal{A}$ to defeat black-box verification in~\cite{sablayrolles2020radioactive}: $\mathcal{A}$ can check if any query to its prediction API matches a sample in $\tilde{D}$ and return, for example, a random response in case of a match. \newtext{As a future work, we will check %Further experiments are needed to ascertain 
if this constitutes a viable approach for $\mathcal{A}$ to evade watermark verification even if the verifier injects some noise into watermarked samples during black-box verification.}
%In this work, we demonstrate that the success of current image dataset watermarking techniques depends on different factors, and they are not robust against \ourmethod that can easily identify and remove watermarked samples from the watermarked dataset. \ourmethod can evade backdoor-based watermarking completely with a small decrease in test accuracy, and evades radioactive watermarking for datasets with small number of classes ($\leq 10$) if the watermarking ratio is less than 30\%. As a future work, we aim to improve \ourmethod using the frequency domain component of images so that it can reverse-engineer the clean version of the watermarked samples and construct a clean dataset that does not contain any watermark. 

%Our work shows that it is still challenging to propose a robust dataset watermarking technique solely based on modifying the pixel or feature space of samples. Potential alternative techniques might be proposed to generate watermarked samples that are closer to the training set in feature domain. However, these methods may be brittle~\citep{radiya2021data}, since adversaries may remove watermarks by using recent model architectures or training methods. We conclude that it may not be probable to design a dataset watermarking technique that remains effective and robust against all future watermark removal attacks or training methods. 

\noindent\textbf{Acknowledgements}: This work was funded in part by research gifts from Intel (via the Private AI Consortium) and Huawei.

%\begin{acks}
%To Robert, for the bagels and explaining CMYK and color spaces.
%\end{acks}

\bibliographystyle{ACM-Reference-Format}
\bibliography{bibliography}


\end{document}